\documentclass[10pt,aps,prb,preprint,twocolumn, fleqn]{revtex4}   

\usepackage{amsmath}    
\usepackage{graphicx}   

\begin{document}
\newcommand{\ab}{\v{a}} 
\newcommand{\ai}{\^{a}} 
\newcommand{\ib}{\^{\i}} 
\newcommand{\tb}{\c{t}} 
\newcommand{\st}{\c{s}}
\newcommand{\Ab}{\v{A}} 
\newcommand{\Ai}{\^{A}} 
\newcommand{\Ib}{\^{I}} 
\newcommand{\Tb}{\c{T}}
\newcommand{\St}{\c{S}}

\newcommand{\muv}{\boldsymbol{\mu}}
\newcommand{\mc}{{\mathcal M}}
\newcommand{\pc}{{\mathcal P}}
 \newcommand{\mv}{\boldsymbol{m}}
\newcommand{\pv}{\boldsymbol{p}}
\newcommand{\tv}{\boldsymbol{t}}
\def\msf{\hbox{{\sf M}}}
\def\msft{\boldsymbol{{\sf M}}}
\def\psf{\hbox{{\sf P}}}
\def\psft{\boldsymbol{{\sf P}}}
\def\Nsf{\hbox{{\sf N}}}
\def\Nsft{\boldsymbol{{\sf N}}}
\def\Tsf{\hbox{{\sf T}}}
\def\Tsft{\boldsymbol{{\sf T}}}
\def\Asf{\hbox{{\sf A}}}
\def\Asft{\boldsymbol{{\sf A}}}
\def\Bsf{\hbox{{\sf B}}}
\def\Bsft{\boldsymbol{{\sf B}}}
\def\Lsf{\hbox{{\sf L}}}
\def\Lsft{\boldsymbol{{\sf L}}}
\def\Ssf{\hbox{{\sf S}}}
\def\Ssft{\boldsymbol{{\sf S}}}
\def\Mtens{\bi{M}}
\def\msfsim{\boldsymbol{{\sf M}}_{\scriptstyle\rm(sym)}}
\newcommand{\mcsim}{ {\sf M}_{ {\scriptstyle \rm {(sym)} } i_1\dots i_n}}
\newcommand{\mcs}{ {\sf M}_{ {\scriptstyle \rm {(sym)} } i_1i_2i_3}}

\newcommand{\beqan}{\begin{eqnarray*}}
\newcommand{\eeqan}{\end{eqnarray*}}
\newcommand{\beqa}{\begin{eqnarray}}
\newcommand{\eeqa}{\end{eqnarray}}

 \newcommand{\suml}{\sum\limits}
 \newcommand{\sumd}{\suml_{\mathcal D}}
\newcommand{\intl}{\int\limits}
\newcommand{\rvec}{\boldsymbol{r}}
\newcommand{\xvec}{\boldsymbol{x}}
\newcommand{\xivec}{\boldsymbol{\xi}}
\newcommand{\Avec}{\boldsymbol{A}}
\newcommand{\Rvec}{\boldsymbol{R}}
\newcommand{\Evec}{\boldsymbol{E}}
\newcommand{\Bvec}{\boldsymbol{B}}
\newcommand{\Svec}{\boldsymbol{S}}
\newcommand{\avec}{\boldsymbol{a}}
\newcommand{\nablav}{\boldsymbol{\nabla}}
\newcommand{\nuvec}{\boldsymbol{\nu}}
\newcommand{\bvec}{\boldsymbol{\beta}}
\newcommand{\vvec}{\boldsymbol{v}}
\newcommand{\jvec}{\boldsymbol{J}}
\newcommand{\nvec}{\boldsymbol{n}}
\newcommand{\pvec}{\boldsymbol{p}}
\newcommand{\mvec}{\boldsymbol{m}}
\newcommand{\evec}{\boldsymbol{e}}
\newcommand{\eps}{\varepsilon}
\newcommand{\la}{\lambda}
\newcommand{\rad}{\mbox{\footnotesize rad}}
\newcommand{\scr}{\scriptstyle}
\newcommand{\latens}{\boldsymbol{\Lambda}}
\newcommand{\pitens}{\boldsymbol{\Pi}}
\newcommand{\cm}{{\cal M}}
\newcommand{\cp}{{\cal P}}
\newcommand{\beq}{\begin{equation}}
\newcommand{\eeq}{\end{equation}}
\newcommand{\ptens}{\boldsymbol{\sf{P}}}
\newcommand{\Ptens}{\boldsymbol{P}}
\newcommand{\Ttens}{\boldsymbol{\sf{T}}}
\newcommand{\Ntens}{\boldsymbol{\sf{N}}}
\newcommand{\Ncal}{\boldsymbol{{\cal N}}}
\newcommand{\Atens}{\boldsymbol{\sf{A}}}
\newcommand{\Btens}{\boldsymbol{\sf{B}}}
\newcommand{\dom}{\mathcal{D}}
\newcommand{\al}{\alpha}
\newcommand{\sym}{\scriptstyle \rm{(sym)}}
\newcommand{\Tcal}{\boldsymbol{{\mathcal T}}}
\newcommand{\Nmc}{{\mathcal N}}
\renewcommand{\d}{\partial}
\def\rmi{{\rm i}}
\def\rme{\hbox{\rm e}}
\def\rmd{\hbox{\rm d}}
\newcommand{\ct}{\mbox{\Huge{.}}}
\newcommand{\Laop}{\boldsymbol{\Lambda}}
\newcommand{\Ssfs}{{\scriptstyle \Ssft^{(n)}}}
\newcommand{\Lsfs}{{\scriptstyle \Lsft^{(n)}}}
\newcommand{\psfr}{\widetilde{\psf}}
\newcommand{\msfr}{\widetilde{\msf}}
\newcommand{\msftr}{\widetilde{\msft}}
\newcommand{\psftr}{\widetilde{\psft}}
\newcommand{\qdot}{\stackrel{\cdot\cdot\cdot\cdot}}
\newcommand{\tdot}{\stackrel{\cdot\cdot\cdot}}
\newcommand{\eref}{(\ref}
\newcommand{\bsy}{\boldsymbol}
\newcommand{\dotj}{\boldsymbol{\dot{J}}}
\newcommand{\psfs}{{\sf P}}
\newcommand{\msfs}{{\sf M}}
\newcommand{\Fvec}{\boldsymbol{F}}
\newcommand{\Qvec}{\boldsymbol{Q}}

\title{Why Jefimenko equations ? (A comment on ``Multipole radiation fields from the Jefimenko equation...'', by R. de Melo e Souza et al, Am. J. Phys. {\bf 77}(1), 67-72 (2009))}
\author{C. Vrejoiu and R. Zus}
 \affiliation{University of Bucaharest, Department of Physics, Bucharest, Romania} 
 \email{vrejoiu@fizica.unibuc.ro;roxana.zus@fizica.unibuc.ro}   
\date{\today}
\begin{abstract}
In this comment, we discuss some features of the multipolar expansion of the power radiated by a confined system of charges and currents, and the possibility of generalization to a higher arbitrary order of the multipolar expansion. 
\end{abstract}

\maketitle

This comment does not intend only to criticize Ref. \cite{Melo} which, generally,  contains correct results, but to make the interested reader aware of a certain type of problems around existing results and on the fact that they are still open to new theoretical and pedagogical contributions. \par Usually, for obtaining the multipolar expansion of the electromagnetic field, one deals with the integral on a finite domain $\dom$ of a function $F(\xvec',\Rvec)$ having the suport included in $\dom$:
\beqa\label{F0}
{\mathcal F}(\xvec)=\intl_{\dom}F(\xvec',\Rvec)\,\rmd^3x'.
\eeqa
Here $\Rvec=\xvec-\xvec'$, the point $P(\rvec)$ is in the exterior of $\dom$, and the origin of the Cartesian coordinates is $O\in \dom$. Denoting by $d$ the linear dimension of $\dom$, one obtains an ${\mathcal F}$  series expansion in powers of $d/r$, $r=\vert\xvec\vert$, by the following procedure:
\beqa\label{F1}
 &~&{\mathcal F}(\xvec)=\intl_{\dom}\left[F(\xivec,\Rvec)\right]_{\xivec=\xvec'}\,\rmd^3x'\nonumber\\
&=&\intl_{\dom}\big[\suml_{n\ge 0}\frac{(-1)^n}{n!}x'_{i_1}\,\dots\,x'_{i_n}\,\d_{i_1}\,\dots\,\d_{i_n}F(\xivec,\xvec)\big]_{\xivec=\xvec'}\rmd^3x'\nonumber\\
&=&\intl_{\dom}\suml_{n\ge 0}\frac{(-1)^n}{n!}x'_{i_1}\,\dots\,x'_{i_n}\,\d_{i_1}\,\dots\,\d_{i_n}F(\xvec',\xvec)\,\rmd^3x'.
\eeqa
With $F$ supposed satisfying all necessary conditions, we can invert the derivative, summation and integration operations such that
\beqa\label{F2}
{\mathcal F}(\xvec)
=\suml_{n\ge 0}\frac{(-1)^n}{n!}\,\d_{i_1}\dots\d_{i_n}\intl_{\dom}x'_{i_1}\dots x'_{i_n}F(\xvec',\xvec)\,\rmd^3x'.
\eeqa 
Equation \eref{F2}) is just the basic formula for providing the multipolar expansions in Cartesian coordinates.
\par Equations (16) and (12) from \cite{Melo} are Jefimenko equations and can be written in the following form:
\beqa\label{JB}
\Bvec(\xvec,t)&=&\frac{1}{c}\intl\nablav\times\frac{[\jvec]}{R}\,\rmd^3x',\nonumber\\
\Evec(\xvec,t)&=&-\int\nablav\frac{[\rho]}{R}\rmd^3x'-\frac{1}{c^2}\int\frac{\dot{[\jvec]}}{R}\,\rmd^3x'.
\eeqa
These  equations, in fact, introduce the retarded potentials $\Avec$ and $\Phi$ if it is possible to invert the order of the derivative and integral operations. In \cite{Melo} this possibility is not considered. Consequently, the multipole expansion of the field should be  calculated using the Taylor series of the integrand from Jefimenko equations. Just in such circumstances Jefimenko equations would be usefull. But, in this case, it is not possible to perform the multipole expansion of the field in an arbitary point from the exterior of the domain $\dom$. Otherwise, the calculation of the multipole expansions is the same as in the case of using the potentials, but with some unnecessary complications as the permanent specification of some derivatives expressing the relation between fields and potentials. In the case of radiation field, the problem of multipole expansion is simplified in this respect. 
\par The radiation field is defined in Ref. \cite{Melo} by retaining in the expressions of the fields $\Evec$ and $\Bvec$ only the terms of order $1/r$. However, we point out the insufficiency of this condition when calculating the radiated angular momentum. In this case, the terms of order $1/r^2$ are also necessary. Equations (28) and (29) from Ref. \cite{Melo}, obtained from a relative complicated calculation, are, in fact, well known from Ref. \cite{Landau} (equations (66.3)). A direct verification of these equations, suggested  in a footnote on page 184 from Ref. \cite{Landau}, is given in Ref. \cite{cvedp}, and also in Ref.\cite{cvdn-arx}. Proving the equations (66.3) from Ref. \cite{Landau}, one uses, in fact, Jefimenko's equations since  equations $\Bvec=\nablav\times\Avec$ and $\Evec=-\nablav\Phi-\frac{1}{c}\dot{\Avec}$ are written  by inverting the order of the  $\nabla$ and  the integration operations, giving the retarded potentials $\Avec$ and $\Phi$. Therefore, equations (28) and (29) from Ref. \cite{Melo} correspond, as result as well as demonstration, to equations (66.3) from Ref. \cite{Landau} which are written in the following form:
\beqa\label{66.3}
\Bvec_{\scriptstyle rad}(\xvec,t)&=&\frac{1}{c}\dot{\Avec}_{\scriptstyle rad}\times\nuvec,\nonumber\\
\Evec_{\scriptstyle rad}(\xvec,t)&=&\frac{1}{c}(\dot{\Avec}_{\scriptstyle rad}\times \nuvec)\times\nuvec.
\eeqa
Here, $\nuvec=\xvec/r$ and
$\Avec_{\scriptstyle rad}$ is  the radiation vector potential  given by equation (66.2) from Ref. \cite{Landau}:
\beqa\label{66.2}
\Avec_{\scriptstyle rad}(\xvec,t)=\frac{1}{cr}\intl_{\dom}\jvec(\xvec',\,t_0+\frac{1}{c}\nuvec\cdot\xvec')\,\rmd^3x',
\eeqa
where $t_0=t-r/c$. This last expression is present in equations  (28) and (29) from Ref. \cite{Melo}.
\par Another issue we want to emphasize in the present comment refers to the calculation technique used for multipolar expansion in Ref. \cite{Melo}. As far as we know, there are several results in the systematics of multipolar expansions, results not used by the authors of Ref. \cite{Melo}. The situation is similar in Ref. \cite{Raab}, a publication dedicated to this problem, comprising valuable results, but with a calculation technique equally primitive and slow. An old, short and self-consistent article published in 1978 in Am.J.Phys. (Ref. \cite{Castell}) may be used to introduce, without additional computational complications, the general expression for the multipolar expansion of the electromagnetic field. One needs just a minimal set of notions on tensor calculus and to be able to choose some simple notation for tensor contraction.
 
\par Let be the Taylor series of the integrand from equation \eref{66.2})  considering only the dependence of $\xvec'$ through the retarded time. By introducing the arbitrary vector $\xivec$, we can write 
\beqan
&~&\jvec(\xivec,t_0+\frac{1}{c}\nuvec\cdot\xvec')\\
&=&\suml_{n\ge 0}\frac{1}{n!}x'_{i_1}\dots x'_{i_n}\,\big[\d'_{i_1}\dots\d'_{i_n}\jvec(\xivec,t_0+\frac{1}{c}\nuvec\cdot\xvec')\big]_{\xvec'=0}.
\eeqan
Because 
\beqan
\d'_i\jvec(\xivec,t_0+\frac{1}{c}\nuvec\cdot\xvec')=\frac{1}{c}\nu_i\dot{\jvec}(\xivec,t_0+\frac{1}{c}\nuvec\cdot\xvec'), 
\eeqan
we can write

\beqa\label{dezvJ}
&~&\jvec(\xivec,t_0+\frac{1}{c}\nuvec\cdot\xvec')\nonumber\\
&=&\sum_{n\ge 0}\frac{1}{c^nn!}\nu_{i_1}\dots \nu_{i_n}x'_{i_1}\dots x'_{i_n}
\frac{\d^n}{\d t^n}\jvec(\xivec,t_0)\nonumber\\
&=&\sum_{n\ge 0}\frac{1}{c^nn!}(\nuvec\cdot\xvec')^n\frac{\d^n\jvec(\xivec,t_0)}{\d t^n}.
\eeqa

\par Starting from this expansion and applying the procedure introduced in Ref. \cite{Castell} for the static case and then generalized to the radiation field in Ref. \cite{cvdn-arx}, one obtains the general form of the expansion:
\beqa\label{d1A}
\Avec_{\scriptstyle rad}(\xvec,t)=\frac{1}{cr}\intl_{\dom}\suml_{n\ge 0}\frac{1}{n!c^n}(\nuvec\cdot\xvec')^n\frac{\d^n\jvec(\xvec',t_0)}{\d t^n}\,\rmd^3x'.
\eeqa
Let us define
\beqa\label{a^n}
\bsy{a}^{(n)}&=&\intl_{\dom}(\nuvec\cdot\xvec')^n\,\jvec(\xvec',t_0)\,\rmd^3x' \nonumber\\
&=&\evec_i\nu_{i_1}\dots\nu_{i_n}\intl_{\dom}x'_{i_1}\dots x'_{i_n}\,J_i(\xvec',t_0)\,\rmd^3x', 
\eeqa
$\evec_i$ being the unit vectors of the Cartesian axes. Introducing the consequence of the continuity equation
\beqan
J_i(\xvec',t_0)=\nablav'\cdot[x'_i\jvec(\xvec',t_0)]+x'_i\frac{\d\rho(\xvec',t_0)}{\d t} ,
\eeqan
we obtain
\beqa\label{a^n-2}
\bsy{a}^{(n)}&=&\evec_i\nu_{i_1}\dots\nu_{i_n}\intl_{\dom}x'_{i_1}\dots x'_{i_n}\nablav'\cdot[x'_i\jvec(\xvec',t_0)]\,\rmd^3x'\nonumber\\
&+&\evec_i\nu_{i_1}\dots\nu_{i_n}\intl_{\dom}x'_{i_1}\dots x'_{i_n}x'_i\frac{\d\rho(\xvec',t_0)}{\d t}\,\rmd^3x' .
\eeqa
Let us introduce the $n-th$ order electric moment of the charge distribution by the components
\beqa\label{e-mom}
P_{i_1\dots i_n}(t)=\intl_{\dom}x'_{i_1}\dots x'_{i_n}\,\rho(\xvec',t)\,\rmd^3x',
\eeqa
and the associated vector
\beqa\label{Pvec}
\bsy{\mathcal P}(\nuvec,t;n)=\nu_{i_1}\dots\nu_{i_{n-1}}P_{i_1\dots i_{n-1}\,i}(t)\,\evec_i
\eeqa
such that
\beqan
\bsy{a}^{(n)}&=&-\evec_i\nu_{i_1}\dots\nu_{i_n}\intl_{\dom}x'_i\jvec(\xvec',t_0)\cdot\nablav'(x'_{i_1}\dots x'_{i_n})\,\rmd^3x'\\
&+&\dot{\bsy{\mathcal P}}(\nuvec,t_0;n+1)
\eeqan
where the cancellation of a surface integral is considered. Because  of $\nablav'x'_i=\evec_i$ and of the symmetry of the contracted indices, we can write
\beqan
\!\!\!\!\!&~&\bsy{a}^{(n)}=-n\evec_i\nu_{i_1}\dots\nu_{i_n}\intl_{\dom}x'_{i_1}\dots x'_{i_{n-1}}x'_iJ_{i_n}(\xvec',t_0)\,\rmd^3x'\\
\!\!\!\!\!&+&\dot{\bsy{\mathcal P}}(\nuvec,t_0;n+1)=-n\bsy{a}^{(n)}-n\evec_i\nu_{i_1}\dots\nu_{i_n}\\
\!\!\!\!\!&~&\intl_{\dom}x'_{i_1}\dots x'_{i_{n-1}}[x'_iJ_{i_n}(\xvec',t_0)-x'_{i_n}J_i(\xvec',t_0)]\,\rmd^3x'\\
\!\!\!\!\!&+& \dot{\bsy{\mathcal P}}(\nuvec,t_0;n+1).
\eeqan
So, we can write
\beqan
&~&\bsy{a}^{(n)}=-\frac{n}{n+1}\,\evec_i\nu_{i_1}\dots\nu_{i_n}\\
&~&\intl_{\dom}x'_{i_1}\dots x'_{i_{n-1}}[x'_iJ_{i_n}(\xvec',t_0)-x'_{i_n}J_i(\xvec',t_0)]\,\rmd^3x'\\
&~&+\frac{1}{n+1}\dot{\bsy{\mathcal P}}(\nuvec,t_0;n+1)\\
&~&=-\frac{n}{n+1}\,\evec_i\,\eps_{ii_nk}\nu_{i_n}\,\nu_{i_1}\dots\nu_{i_{n-1}}\intl_{\dom}x'_{i_1}\dots x'_{i_{n-1}}\\
&~&[\xvec'\times\jvec(\xvec',t_0)]_k\,\rmd^3x'+\frac{1}{n+1}\dot{\bsy{\mathcal P}}(\nuvec,t_0;n+1)
\eeqan
We introduce, as in Ref. \cite{Castell}, the $n-th$ order magnetic moment by the Cartesian components:
\beqa\label{magnmom}
\!\!\!\!\!\!\!\!\!\!\!\!\!\!\!&~&M_{i_1\dots i_n}(t)=\frac{n}{(n+1)c}\intl_{\dom}x'_{i_1}\dots x'_{i_{n-1}}[\xvec'\times\jvec(\xvec',t)]_{i_n}\,\rmd^3x',\nonumber\\
\!\!\!\!\!&~&
\eeqa
and the associated vector
\beqa\label{Mvec}
\bsy{\mathcal M}(\nuvec,t;n)=\nu_{i_1}\dots\nu_{i_{n-1}}\,M_{i_1\dots i_{n-1}\,i}\,\evec_i .
\eeqa
Therefore, we can write
\beqan
\bsy{a}^{(n)}&=&-c\,\evec_i\eps_{ii_nk}\nu_{i_n}{\mathcal M}_k(\nuvec,t_0;n)+\frac{1}{n+1}\dot{\bsy{\mathcal P}}(\nuvec,t_0;n+1)\\
&=&c\,\bsy{\mathcal M}(\nuvec,t_0;n)\times\nuvec+\frac{1}{n+1}\dot{\bsy{\mathcal P}}(\nuvec,t_0;n+1) .
\eeqan
With the last result, equation \eref{d1A}) can be written as
\beqan
\!\!\!\!\!\!\!\!\!\!&~&\Avec_{\scriptstyle rad}(\xvec,t)=\frac{1}{r}\suml_{n\ge 1}\frac{1}{n!c^n}\frac{\rmd^n}{\rmd t^n}\bsy{\mathcal M}(\nuvec,t_0;n)\times\nuvec\\
\!\!\!\!\!\!\!\!\!\!&+&\frac{1}{r}\suml_{n\ge 0}\frac{1}{(n+1)!c^{n+1}}\frac{\rmd^{n+1}}{\rmd t^{n+1}}\bsy{\mathcal P}(\nuvec,t_0;n+1) .
\eeqan
After performing a change in the summation index of the last sum,the expansion of the vector potential can be written as
\beqa\label{darad}
\Avec_{\scriptstyle rad}(\xvec,t)&=&\frac{1}{r}\suml_{n\ge 1}\frac{1}{n!c^n}\frac{\rmd^n}{\rmd t^n}\big[
\bsy{\mathcal M}(\nuvec,t_0;n)\times\nuvec\nonumber\\
&+& \bsy{\mathcal P}(\nuvec,t_0;n)\big]
\eeqa

Using equations \eref{66.3}) and \eref{darad}), we obtain
\beqa\label{deB}
&~&\Bvec_{\scriptstyle rad}(\xvec,t)\\
&=&\frac{1}{cr}\suml_{n\ge 1}\frac{1}{c^nn!}\frac{\rmd^{n+1}}{\rmd t^{n+1}}
\left[\bsy{\mathcal M}(\nuvec,t_0;n)\times\nuvec+
\bsy{\mathcal P}(\nuvec,t_0;n)\right]\times \nuvec.\nonumber
\eeqa

Equation \eref{deB}) is sufficient for calculating the total power radiated by the given charge and current system, but this calculation is possible in a general form if instead of the ``primitive'' tensors $\msft^{(n)}$ and $\psft^{(n)}$ one introduces total symmetric and trace free ({\bf STF}) tensors. In this respect there are some results in literature (see Refs. \cite{Thorne},\cite{Damour}, \cite{cvdn-JPA}).
\par A last observation regarding Ref. \cite{Melo}: equation (48) is a correct, but not usefull result. Instead of $\bsy{Q}(\xivec,t)$ given by the mentioned equation, one can define 
\beqan
\bsy{Q}(\xivec,t)=\intl_{\dom}[3(\xivec\cdot\xvec')\xvec'-r^{'2}\xivec]\,\rho(\xvec',t)\,\rmd^3x'
\eeqan
which correspods to the {\bf STF} tensor $\widetilde{\psft}^{(2)}$
\beqan
\widetilde{\psfs}_{ij}=\intl_{\dom}(3x'_ix'_j-r^{'2}\delta_{ij})\,\rho(\xvec',t)\,\rmd^3x'
\eeqan
introduced in the field expansion instead of the tensor representing the electric 4-polar moment $\psft^{(2)}$. 
\par We point out that field invariance to the substitution $\psft^{(2)}\,\rightarrow\,\widetilde{\psft}^{(2)}$ is an isolated case. For other substitutions of $\psft^{(n)}$ or $\msft^{(n)}$ by their projections on the space of {\bf STF} tensors the field is not invariant and, consequently, the physical results also not. This circumstance is pointed out in Ref. \cite{Raab}, too. However, these substitutions are possible with the price of introducing modified {\bf STF} tensors generally different from the projections of the ``primitive'' tensors \cite{cvdn-JPA}.

\begin{acknowledgments}
The work of RZ was supported by grant ID946 (no.\ 44/2007) of Romanian National Authority for Scientific Research.
\end{acknowledgments}


\begin{thebibliography}{5}
\bibitem{Melo}
R. de Melo e Souza, M. V. Cougo-Pinto, and C. Farina, ``Multipole radiation fields from Jefimenko equation for the magnetic field and the Panofsky-Phillips equation for the electric field'',Am. J. Phys. {\bf 77}, 67-72 (2009)
\bibitem{Landau}
L. D. Landau  and E. M. Lifchitz,  {\it The Classical Theory of Fields}, (Butterworth-Heinemann Elsevier Science, 2003) Chap. 9, p 184
\bibitem{cvedp}
C. Vrejoiu,  {\it Electrodynamics and Relativity Theory} ( in Romanian), (E. D. P. Bucharest, 1993)
\bibitem{cvdn-arx}
C. Vrejoiu and D. Nicmoru\c{s}, ``Expressing the power radiated by electric charged systems'', arXiv:physics/0307113 (2003)
\bibitem{cvdn-JPA}
C. Vrejoiu and D. Nicmoru\c{s}, ``On the multipole electromagnetic radiation'', J. Phys. A:Math.Gen. {\bf 37}, 1-14 (2004)
\bibitem{Raab}
R. E. Raab, O. L. De Lange, {\it Multipole Theory in Electromagnetism}, (Clarendon Press Oxford, 2005)
\bibitem{Castell}
A. Castellanos, M. Panizo, and J. Rivas, ``Magnetostatic multipoles in Cartesian coordinates'', Am. J. Phys, {\bf 46}, 1116 (1978)
\bibitem{Thorne}
K. S. Thorne, ``Multipole expansions of gravitational radiation'', Rev. Mod. Phys. {\bf 52}, 299 (1980)
\bibitem{Damour}
T. Damour, B. R. Iyer, ``Multipole analysis for electromagnetism and liniarized gravity with irreducible Cartesian tensors'', Phys.Rev. D {\bf 43}, 3259-3272 (1991) 

\end{thebibliography}
\end{document}